\documentstyle[12pt]{article}
\begin{document}
\newcommand{\beq}{\begin{equation}}
\newcommand{\eeq}{\end{equation}}
\newcommand{\beqn}{\begin{eqnarray}}
\newcommand{\eeqn}{\end{eqnarray}}
\newcommand{\slp}{\raise.15ex\hbox{$/$}\kern-.57em\hbox{$\partial
$}}
\newcommand{\slA}{\raise.15ex\hbox{$/$}\kern-.57em\hbox{$A$}}
\newcommand{\lnA}{\raise.15ex\hbox{$/$}\kern-.57em\hbox{$A$}}
\newcommand{\lnC}{\raise.15ex\hbox{$/$}\kern-.57em\hbox{$C$}}
\newcommand{\slB}{\raise.15ex\hbox{$/$}\kern-.57em\hbox{$B$}}
\newcommand{\bP}{\bar{\Psi}}
\newcommand{\bC}{\bar{\chi}}
 \newcommand{\hs}{\hspace*{0.6cm}}

\title{Bosonic description of a Tomonaga-Luttinger model with impurities}
\author{
Victoria Fern\'andez$^{a}$, Kang Li$^{a,b}$ and Carlos Na\'on$^{a}$}
\date{December 1998}
\maketitle

\def\thepage{\protect\raisebox{0ex}{\ } La Plata-Th 98/22}
\thispagestyle{headings}
\markright{\thepage}

\begin{abstract}

\hs We extend a recently proposed non-local version of Coleman's equivalence
between the Thirring and sine-Gordon models to the case in which the
original fermion fields interact with fixed impurities.
We explain how our results can be used in the context of one-dimensional
strongly correlated systems (the so called Tomonaga-Luttinger model) to study the dependence of the charge-density
oscillations on the range of the fermionic interactions.

\end{abstract}

\vspace{3cm}
Pacs: 05.30.Fk\\
\hspace*{1,7 cm}11.10.Lm\\
\hspace*{1,7 cm}71.10.Pm

\noindent --------------------------------

\noindent $^a$ {\footnotesize Depto. de F\'\i sica.  Universidad
Nacional de La Plata.  CC 67, 1900 La Plata, Argentina.\\
E-Mail: naon@venus.fisica.unlp.edu.ar}

\noindent $^b$ {\footnotesize Department of Physics, Zhejiang
University, Hangzhou, 310028, P. R. China.}

\newpage
\pagenumbering{arabic}

\hs In a recent paper \cite{LN}  a {\em non-local}
generalization of Coleman's equivalence between the massive
Thirring and sine-Gordon models \cite{Coleman} was established. Specifically it was
proved the identity of the vacuum to vacuum functionals
corresponding to the following (Euclidean) Lagrangian densities:
\beq {\cal L}{_T}=i\bar{\Psi}\slp\Psi+\frac{1}{2}g^2\int d^2
yJ_{\mu}(x)V_{(\mu )}(x,y) J_{\mu}(y) -m\bP\Psi
\label{1}
\eeq
and
\beq {\cal L}_{SG}
=\frac{1}{2}(\partial_{\mu}\phi(x))^2+\frac{1}{2} \int
d^2y\partial_{\mu}\phi(x)d_{(\mu )}(x,y)\partial_{\mu}\phi(y)
-\frac{\alpha_0}{\beta^2}\cos \beta\phi,
\label{2}
\eeq
where
$J_{\mu}(x)$ is the usual fermion current, $V_{(\mu )}(x,y)$
and $d_{(\mu)}(x,y)$ ($\mu = 0,1$) are arbitrary functions of
$|x-y|$ whose Fourier transforms ($\hat{V}_{(\mu )}(p)$ and
$\hat{d}_{(\mu )}(p)$) must be related by

\beq \frac{1} {\frac{g^2}{\pi}(p_0^2 \hat{V}_{(1)}
+p_1^2\hat{V}_{(0)}) + p^2}= \frac{\beta^2} {4\pi(p^2+
\hat{d}_{(0)} p_0^2+ \hat{d}_{(1)} p_1^2) } \label{3} \eeq For the
sake of clarity, we have omitted the p-dependence of the
potentials. When $V_{(\mu )}(x,y)= \delta^2(x-y)$ and
$d_{(\mu)}(x,y)=0$ one reobtains the usual local equivalence (let
us also mention that with the present convention attractive
potentials correspond to positive values of $V_{(\mu )}$). Of
course, the usual identity \beq m=\frac{\alpha_0}{\beta^2}
\label{4} \eeq also holds in this {\em non-local} context.\\ \hs
The purpose of this letter is to extend the above result in two
directions. First of all we shall consider a coordinate-dependent
mass in our non-local version of the Thirring model. At the same
time we will also add a local interaction between the fermion
current and a classical background field $C_{\mu}$. In other
words, we shall modify equation (\ref{1}) above by setting $m=0$
and adding \beq {\cal L}_{imp}=\bar{\Psi}\lnC\Psi -
m(x)\bar{\Psi}\Psi \label{5} \eeq where the name ${\cal L}_{imp}$
refers to the fact that this density can be used to study the
interaction between electrons and impurities \cite{KF}. Indeed, if
one uses (\ref{1}) (with $m=0$) in order to describe the forward
scattering of one-dimensional spinless electrons \cite{Voit}
\cite{NRT}, then the first (second) term in the rhs of (\ref{5})
models the forward (backward) scattering between electrons and
impurities. In particular, if we choose
\beq C_0(x)=V \delta
(x_{1}-d)= m(x) \label{6} \eeq where V is a constant, together
with \beq C_1(x)=0, \label{7} \eeq our model coincides with the
one recently considered in ref.\cite{YCZC} to study Friedel
charge-density oscillations in a 1d Tomonaga-Luttinger liquid
\cite{TL} with an impurity located at $x_{1}=d$. Moreover, the
bosonic vacuum to vacuum functional we shall derive here can also
be used to compute the 2-point density function and the electron
propagator as functionals of the electron-electron potential. The
former was computed in ref.\cite{Schulz} (for a Coulomb potential
and without impurities) where a Wigner crystal structure was
revealed. Apart from academic interest, these connections are the
main practical motivation for the present computation. Before
making contact with this solid-state application, we now start by
considering the general case ($C_{\mu}(x)$ and $m(x)$ arbitrary).
Using a convenient representation of the functional delta and
introducing the vector field $A_{\mu}$ (Please see ref. \cite{LN}
for details), the partition function of the model under
consideration can be written as \beq Z =  \int DA_{\mu} det(i \slp
+ g \lnA + \lnC -m(x)) e^{-S[A]}, \label{8} \eeq \noindent with
\beq S(A)=\frac{1}{2}\int d^2xd^2y
V_{(\mu)}^{-1}(x,y)A_{\mu}(x)A_{\mu}(y), \label{9} \eeq \noindent
where $V_{(\mu)}^{-1}$ is such that \beq \int d^2z~
V_{(\mu)}^{-1}(z,x) V_{(\mu)}(y,z) = \delta^2 (x-y). \eeq

As it is known, the massive-like determinant in equation (\ref{8}) cannot be exactly solved, even in the local case. However, we were able to write the vacuum to vacuum functional in such a way that non-local terms are not present in the determinant. Then, following \cite{LN}, we can decouple $A_{\mu}$ and $C_{\mu}$ from fermions by performing chiral and gauge transformations in the fermionic path-integral measure, with parameters $\Phi(x)$ and $\eta(x)$ respectively:
\beq
\Psi(x) = exp[-g(\gamma_5 \Phi(x) + i \eta(x))] \chi(x),
\label{10}
\eeq
\noindent (and a similar expression for $\bP(x)$) and writing
\beq
A_{\mu}(x)=\epsilon_{\mu\nu}\partial_{\nu}\Phi (x)-
\partial_{\mu}\eta (x)- \frac{1}{g}C_{\mu}(x).
\label{11}
\eeq
\noindent Taking into account the non-trivial Fujikawa Jacobian \cite{Fuji} associated to the fermionic transformation, which gives a local kinetic term for $\Phi (x)$, we obtain
\beq
Z= \int D\bar\chi D\chi D\Phi D\eta e^{-S_{eff}}
\label{12}
\eeq
\noindent where
\beq
S_{eff}=S_{0B}+ \int d^2x (\bar\chi i\slp\chi - m(x)\bar\chi e^{-2g\gamma_5\Phi}\chi),
\label{13}
\eeq

\beq
\begin{array}{cl}
 S_{0B}=&\frac{g^2}{2\pi}
           \int d^2x (\partial_{\mu}\Phi)^2 \nonumber\\
    ~   &+ \frac{1}{2}\int d^2x d^2y \epsilon_{\mu\lambda}\epsilon_{\mu\sigma}
        V_{(\mu)}^{-1}(y,x) \partial_{\lambda} \Phi (x)
        \partial_{\sigma} \Phi (y) \nonumber\\
     ~   &+\frac{1}{2}\int d^2x d^2y V_{(\mu)}^{-1}(y,x) \partial_{\mu}
        \eta (x)
        \partial_{\mu}\eta (y) \nonumber\\
      ~  &-\int d^2x d^2y [V_{(0)}^{-1}(y,x) \partial_0 \eta (x)
        \partial_1 \Phi (y) \nonumber\\
       ~ &- V_{(1)}^{-1}(y,x) \partial_1 \eta (x)
        \partial_0 \Phi (y)] \nonumber\\
       ~ &- \frac{1}{g}\int d^2x d^2y V_{(\mu)}^{-1}(x,y)
        C_{\mu}(y)(\epsilon_{\mu\sigma}
        \partial_{\sigma}\Phi (x)-\partial_{\mu} \eta (x))\nonumber\\
             ~   &+ \frac{1}{2g^2}\int d^2x d^2y V_{(\mu)}^{-1}(x,y)
        C_{\mu}(x)C_{\mu}(y).
\end{array}
\label{14}
\eeq

\noindent Note that the last term of $S_{OB}$ is field independent (remember that $C_{\mu}(x)$ is a classical function), thus
its contribution can be absorbed in a path-integral normalization constant $N[C_{\mu}]$ which is relevant if one is interested in functional derivatives of $Z$ with respect to $C_{\mu}(x)$. Since here we are mainly concerned with vacuum to vacuum functionals we shall disregard this factor. Its contribution will be recovered at the end of this work, where the evaluation of the fermion current v.e.v. will be sketched. Let us also mention that one should be more careful with this normalization when considering the finite temperature version of the present calculation, which could be easily done by following the lines of ref. \cite{MNT}. On the other hand, the first term in (\ref{14}) comes from the contribution of the previously mentioned fermionic Jacobian.
Exactly as it was done in \cite{LN}, the partition function
for the model  can be formally expanded in powers of $m(x)$. In fact, the x dependence of this perturbative parameter, together with the appearance of $C_{\mu}(x)$ in the bosonic action are the new features of the present computation. As far as these functions are well-behaved one can assume the existence of every term in the following series:

\beq
Z=\sum_{n=0}^{\infty}\frac{1}{n!}<\prod_{j=1}^n \int d^2 x_j
m(x_j)\bar\chi (x_j) e^{-2g\gamma_5\Phi (x_j)}\chi (x_j )>_{0}
\label{15}
\eeq

It is not difficult to convince oneself that the same manipulations used in \cite{LN},
in order to compute fermionic and bosonic v.e.v's at every order of perturbation theory,
also work in the present case. This allows to write
\beqn
Z&=\sum_{k=0}^{\infty}\frac{1}{(k!)^2}\int\prod_{i=1}^k d^2x_i
d^2y_i m(x_i) m(y_i)exp\{ -\int \frac{d^2p}{(2\pi)^2}[\frac{2\pi}{p^2}\nonumber\\
&-\frac{2\pi g^2(\hat{V}_{(0)}^{-1}p_0^2+\hat{V}_{(1)}^{-1}p_1^2 )}
{ g^2(\hat{V}_{(0)}^{-1}p_0^2+\hat{V}_{(1)}^{-1}p_1^2)p^2+\pi\hat{V}_{(0)}^{-1}
\hat{V}_{(1)}^{-1}p^4}]D(p,x_i,y_i)D(-p,x_i,y_i)\}
\nonumber\\
&exp\{ -\int\frac{d^2p}{(2\pi)^2}\frac{2i\hat{V_{(\mu)}}^{-1}(-p)\hat{C_{\mu}}(-p)}{\Delta (p)}
(C(p)p_{\mu}-2\epsilon_{\mu\nu}p_{\nu}B(p))D(-p,x_i,y_i) \}
\label{16}
\eeqn
\noindent where, for simplicity, we have gone to momentum space and defined
\beq
     A(p) = \frac{g^2}{2\pi}~ p^2 +
     \frac{1}{2}[\hat{V}_{(0)}^{-1}(p) p_1^2 +
           \hat{V}_{(1)}^{-1}(p) p_0^2],
\label{17}
\eeq

\beq
B(p) = \frac{1}{2}[\hat{V}_{(0)}^{-1}(p) p_0^2 +
           \hat{V}_{(1)}^{-1}(p) p_1^2],
\label{18}
\eeq

\beq
C(p) = [\hat{V}_{(0)}^{-1}(p) - \hat{V}_{(1)}^{-1}(p)] p_0 p_1,
\label{19}
\eeq

\beq
\Delta = C^2(p)-4A(p)B(p),
\label{20}
\eeq
and
\beq
D(p,x_i,y_i) =\sum_i (e^{ipx_i}-e^{ipy_i}).
\label{21}
\eeq
\noindent In these expressions $\hat{\Phi},\hat{\eta}$ and $\hat{V}_{(\mu)}$ are the Fourier
transforms of $\Phi, \eta$ and $V_{(\mu)}$ respectively. Equation
(\ref{16}) is our first non-trivial result. It is the extension of
the result presented in \cite{LN} to the case $m \rightarrow m(x)$
and $C_{\mu}(x)\neq 0$.

Let us now briefly make our second relevant observation which
concerns the way of constructing a purely bosonic model that gives
the same expansion as in (\ref{16}). It is straightforward to show
that such model can be obtained from equation (\ref{2}) by setting
$\alpha_0=0$ and adding \beq {\cal L'}_{imp}=
F_{\mu}(x)\partial_{\mu}\phi (x)-\frac{\alpha_0 (x)}{\beta^2}\cos
\beta\phi (x)\label{22} \eeq where $F_{\mu}(x)$ represents a
couple of classical functions to be related
 to the $C_{\mu}$'s and $\alpha_0 (x)$ is the x-dependent version of
 $\alpha_0$, to be related, of course, to $m(x)$.
 Again, going to momentum space and employing standard procedures to evaluate
each v.e.v., the partition function $Z'$ corresponding to this
generalized sine-Gordon model can be expressed as \beqn
Z'=\sum_{k=0}^{\infty}[\frac{1}{k!}]^2 \int\prod_{i=1}^k d^2x_i
d^2y_i \frac{\alpha (x_i)}{\beta^2}\frac{\alpha
(y_i)}{\beta^2}\nonumber\\ ~exp\{-\frac{\beta^2}{4} \int
\frac{d^2p}{(2\pi)^2}\frac{D(p,x_i,y_i)D(-p,x_i,y_i)}
              {\frac{1}{2}p^2+\frac{1}{2}(\hat{d}_{(0)}(p) p_0^2+
\hat{d}_{(1)}(p) p_1^2)} \}\nonumber\\
exp\{\beta\int
\frac{d^2p}{(2\pi)^2}\frac{\hat{F}_{\mu}(-p)p_{\mu}}
{p^2+p_0^2\hat{d_0}(p)+p_1^2\hat{d_1}(p)}D(-p,x_i,y_i)\}
\label{23}
\eeqn

By comparing equation (\ref{23} ) with equation (\ref{16}), we
find that both series are identical if not only equation (\ref{3})
holds, but one also has:
\beq m(x)=\frac{\alpha_0 (x)}{\beta^2}
\label{24} \eeq and \beq
2i\hat{V}_{(\mu)}^{-1}(-p)\hat{C}_{\mu}(-p)\frac{p_{\mu} C(p)
-2\epsilon_{\mu\nu} p_{\nu}B(p) }{\Delta (p)}=
\beta\frac{\hat{F_{\mu}}(-p)p_{\mu}}{p^2+ \hat{d}_{(0)} p_0^2+
\hat{d}_{(1)} p_1^2} \label{25} \eeq

Therefore, we have obtained an equivalence between the partition
functions $Z$ and $Z'$ corresponding to the  non-local Thirring and
sine-Gordon models with extra interactions defined above, in equations
(\ref{5}) and (\ref{22}) respectively. Note that, apart from eq.(\ref{3})
which was already obtained in \cite{LN} and eq.(\ref{24}) which is a trivial
generalization of eq.(\ref{4}), eq.(\ref{25}) should be considered the
specific original contribution of the present work. As stated in the
introductory paragraph, this path-integral identification allows to make
contact with recent descriptions of 1d electronic systems (the so called Tomonaga-Luttinger model) in the presence of
fixed (not randomly distributed) impurities \cite{YCZC}. We shall devote the
remainder of this letter to briefly discuss this possibility. To be specific we shall consider the mean value of the fermionic current defined by:
\beq
 \langle j_{\mu} \rangle = -\frac{1}{Z} \frac{\delta Z}{\delta  C_{\mu}}
\label{26}
\eeq
This is an interesting object because $j_{0}$ is the charge density whereas $j_{1}$ is the electric current.
Please recall that the usual version of the Tomonaga-Luttinger model, where only density-density fluctuations are taken into account, corresponds to the choice \cite{LN}
\beq
\hat{V}_{(1)}=0
\label{27}
\eeq
Concerning the impurity, it is evident that after performing the functional derivative in eq.(\ref{26}) one has to use equations (\ref{6}) and (\ref{7}) for $C_{0}$ and $C_{1}$. At this point the way of exploiting the above depicted procedure, i.e. the equivalence between $Z$ and $Z'$, as a bosonizing scheme, becomes apparent. Indeed, one has just to enforce condition (\ref{27}) in (\ref{3}) and (\ref{25}) and then use eq.(\ref{26}) with $Z'$ instead of $Z$. For the most relevant case of $j_{0}$ one has to keep in mind that there will be an extra term coming from the disregarded $C_{0}$-dependent factor in $Z$ (Please remember the comments on $N[C_{\mu}]$ that followed equation (\ref{14})). However this is not a big trouble, since the corresponding contribution can be obtained by inspection yielding
\beq
\langle j_{0}(x) \rangle = -\frac{1}{Z'} \frac{\delta Z'}{\delta  C_{0}(x)}+ \frac{1}{g^2}V V_{(0)}^{-1}(x_{1}-d)
\label{28}
\eeq
We want to stress that this expression gives the charge density of a Tomonaga-Luttinger electronic liquid, with a non-magnetic impurity located at $x_{1}=d$, as functional of the electron-electron potential. Thus our present proposal can be considered as an alternative path-integral approach to explore the joint effect of impurities and potentials on the charge density behavior.
The explicit calculation of the first term would allow to check the result recently reported in ref.\cite{YCZC}, where a strong dependence of the density oscillations on the range of the interaction was obtained . We hope to consider this issue in a forthcoming article \cite{next}.

\vspace{1cm}

\newpage
{\bf Acknowledgments}

K.L. thanks  Prof. R.Gamboa Sarav\'{\i} for
hospitality in Universidad Nacional de La Plata (UNLP), Argentina.\\
V.F. and C.N. are partially supported by Universidad Nacional de La Plata (UNLP) and Consejo Nacional de Investigaciones
Cient\'{\i}ficas y T\'ecnicas (CONICET), Argentina. The authors also recognize the support of the Third World Academy of Sciences (TWAS).\\

\end{document}